\documentclass[journal,twoside,web]{arxivcolor} 
\usepackage{tmi}
\usepackage{cite}
\usepackage{amsmath,amssymb,amsfonts}
\usepackage{algorithmic}
\usepackage{graphicx}
\usepackage{tabularx}
\usepackage{multirow}
\usepackage{booktabs}
\usepackage{textcomp}
\usepackage{float}
\def\BibTeX{{\rm B\kern-.05em{\sc i\kern-.025em b}\kern-.08em
    T\kern-.1667em\lower.7ex\hbox{E}\kern-.125emX}}
\begin{document}
\title{Bidirectional Mapping Generative Adversarial Networks for Brain MR to PET Synthesis}
\author{Shengye Hu, Baiying Lei, Yong Wang,  Zhiguang Feng,  Yanyan Shen, Shuqiang Wang
\thanks{Corresponding author: Shuqiang Wang, Email:sq.wang@siat.ac.cn}
\thanks{Shengye Hu and Baiying Lei contributed equally to this work.}
\thanks{ Shengye Hu, Yanyan Shen and Shuqiang Wang are with the Shenzhen Institutes of Advanced Technology, Chinese Academy of Sciences, Shenzhen 518055,China}
\thanks{Baiying Lei is with the School of Biomedical Engineering, Shenzhen University, Shenzhen, China}
\thanks{Yong Wang is with School of Information Science and Engineering, Central South University, Changsha 410083, China}
\thanks{Zhiguang Feng is with College of Intelligent Systems Science and Engineering, Harbin Engineering University, 150001 Harbin, China}
}

\maketitle

\begin{abstract}
Fusing multi-modality medical images, such as MR and PET, can provide various anatomical or functional information about human body. But PET data is always unavailable due to different reasons such as cost, radiation, or other limitations. In this paper, we propose a 3D end-to-end synthesis network, called Bidirectional Mapping Generative Adversarial Networks (BMGAN), where image contexts and latent vector are effectively used and jointly optimized for brain MR-to-PET synthesis. Concretely, a bidirectional mapping mechanism is designed to embed the semantic information of PET images into the high-dimensional latent space. And the 3D DenseU-Net generator architecture and the extensive objective functions are further utilized to improve the visual quality of synthetic results. The most appealing part is that the proposed method can synthesize the perceptually realistic PET images while preserving the diverse brain structures of different subjects. Experimental results demonstrate that the performance of the proposed method outperforms other competitive cross-modality synthesis methods in terms of quantitative measures, qualitative displays, and classification evaluation.
\end{abstract}


\section{Introduction}
\label{sec:introduction}
\IEEEPARstart{A}{s} the cornerstone for precision medicine, medical images have become a requisite component of public health studies. It contains various imaging modalities that are used to provide an intuitive insight of the interior of the human body in order to assist the radiologists and clinicians to detect or treat diseases more efficiently\cite{ref_1}. It has been widely-recognized that each type of modalities presents various anatomical or functional information about human body. Consequently, complementary imaging modalities are always acquired simultaneously to indicate the disease areas, present the various tissue properties, and help to make an accurate and early diagnosis. For example, recent studies have reported that magnetic resonance (MR) imaging and positron emission tomography (PET) measurements are among the effective biomarkers for Alzheimer Disease (AD) progression and Mild Cognitive Impairment (MCI) conversion prediction\cite{ref_2,ref_3}. However, patients may not have all demanded imaging modalities due to a variety of reasons. And a typical example is the lack of PET data. Compared to the widely available and non-invasive MR imaging, PET scan is not offered in the majority of medical centers in the world. Moreover, it will increase lifetime cancer risk due to the usage of radioactive tracer. These factors jointly result in the scarcity of PET data.

In fact, the above observations reflect a common dilemma in the clinic analysis of multimodal medical images—some imaging modalities are unavailable or lacking due to different reasons such as cost, radiation, or other limitations. In such cases, medical imaging synthesis is a novel and effective solution. Given a subject’s image in source modality, the cross-modality synthesis of medical images is to accurately estimate the respective image of the same subject in target modality, such as MR-to-PET. Cross-modality synthesis has huge potentials in clinic applications, especially in the scenarios that offers an automatable, low-cost source of highly demanded imaging modality data for clinic diagnosis. In addition, it also can be used as an anonymization tool~\cite{ref_4}, which means the synthetic images can be shared in the public without caring the protection of personal health information.

Cross-domain synthesis of medical images has recently become one of the mainstream research directions in medical imaging. The existing methods are mainly divided into two categories, including registration-based methods and learning-based methods.

Most registration-based methods~\cite{ref_5,ref_6,ref_7} need calculate an atlas according to the co-registered images. But these methods highly relied on the accuracy of registration, and their application is very limited, which suffers from across-subject differences in underlying morphology~\cite{ref_8}. An alternative is to use learning-based methods that learning a non-linear mapping from source image to target image. For instance, Jog et al.~\cite{ref_9}  learned a nonlinear regression with random forest to carry out cross-modality synthesis of high-resolution images from low resolution scans. Huynh et al.~\cite{ref_10}  presented an approach to synthesize CT from MRI using random forest as well. The disadvantages of these traditional machine learning methods are they largely rely on the handcrafted features and require professional knowledge of domain experts, which results in a challenge for non-experts to make researches in this field.

In recent years, deep learning methods have made a major breakthrough in many fundamental computer vision applications, such as image classification~\cite{ref_27} , object detection~\cite{ref_28}, and image caption~\cite{ref_29}. Due to the strong feature extraction and automatic learning capabilities of convolutional neural networks (CNN), it has also achieved excellent performance in the cross-modality synthesis of medical images. To the best of our knowledge, the earliest published literature in this field may be~\cite{ref_11}. This paper used a CNN to learn a nonlinear mapping from MR to Fludeoxyglucose PET (FDG-PET). And Xiang et al.~\cite{ref_12} presented a deep auto-context CNN that generate full-dose PET images from the low-dose PET images. These models usually used the $\mathcal{L}_{1}$ or $\mathcal{L}_{2}$ distance as the loss function, which often results in the blurry estimations of generated images due to the averaged effect. The emergence of generative adversarial networks (GAN)~\cite{ref_13} with outstanding generative capabilities has further encouraged the development of medical imaging synthesis. Ben et al.~\cite{ref_14} presented a novel system combined by FCN and conditional GAN for the generation of virtual PET images using CT scans. Lei et al.~\cite{ref_15} train the deep GAN in a quasi-3D way and simplify the generator, which leads to faster training yet better synthetic quality. Salman et al.~\cite{ref_16} proposed a novel pGAN for paired image synthesis, and it yields visually and quantitatively enhanced accuracy in multi-contrast MR synthesis compared to the state-of-the-art methods.

Although these algorithms achieve remarkable results, they are suffer from the same problem: trivially appending a randomly sampled latent vector is difficult to generate plausible images with the diverse structures, because the generator learns to largely ignore the latent vector input without any prior knowledge in the training process of GANs ~\cite{ref_17,ref_18,ref_19,ref_20}. Ideally, different latent vector input should correspond to the diverse synthetic images. But the generator is easy to map several different latent vectors to the same output in reality, especially for the generation of brain images that have diverse structural details (e.g. gyri and sulci) between different subjects.

The second problem is that most of previous methods are designed for synthesizing individual 2D slices along one of the axial, coronal and sagittal direction. This results in discontinuous estimation and the loss of spatial information, which are harmful for medical image synthesis. Although some recent works~\cite{ref_21,ref_22,ref_23} generate 3D image patches and merge all patches by averaging the overlapped estimation values to obtain the whole 3D estimated image. But learning on imaging patches is insufficient to extract both local and global contextual information among volume voxels, which obviously affects the generative capacity of the deep models.

In this paper, inspired by the appealing success of GANs works~\cite{ref_17, ref_24, ref_25}, we propose a novel 3D end-to-end network, called Bidirectional Mapping GAN (BMGAN), where image contexts and latent vector are effectively used and jointly optimized for the brain MR-to-PET synthesis. The novelties and contributions of this paper are as follows. (i) We proposed a novel end-to-end 3D medical imaging synthesis network with a bidirectional mapping mechanism. As an attempt to bridge the gap between network generative capability and real medical images, the proposed method not only focused on synthesizing perceptually realistic PET images from MR images, but also concentrated on reflecting the diverse brain attributes of different subjects. (ii) An advanced 3D DenseU-Net was adopted as the generator architecture to optimally preserved the spatial intrinsic features of medical images and eliminated the slices discontinuity problem caused by 2D network. Besides, the extensive functions containing the adversarial loss, the KL-divergence constraints, the reconstruction loss, and the perceptual loss  were devised to further improve the visual quality of synthetic images.

The remaining parts of this paper are organized as follows. Section II provides a detailed description of the proposed BMGAN, including the basic ideas, the DenseU-Net generator architecture, and the extensive objective functions. This section supplies the information needed to the repetition of experiments. The extensive experiments and the ablation study of the proposed method are presented in section III, which demonstrate the performance of the BMGAN. Finally, we summarize the results of this paper and discuss possible future directions for medical image synthesis in section IV.

A preliminary version of this work has been presented at a conference~\cite{ref_26}. Herein, we (i) extend our method by replacing the residual paths with the dense connection in the generator. (ii) evaluate and further analyze the contribution of the 3D network architecture and the adversarial training strategy. (iii) evaluate and further analyze the improvement after applying DenseU-Net generator architecture and the extensive objective function. (iv) include additional discussions that are not in the conference publication.

\section{Method}
\subsection{Overview}
\begin{figure*}[!t]
\centerline{\includegraphics[width=2\columnwidth]{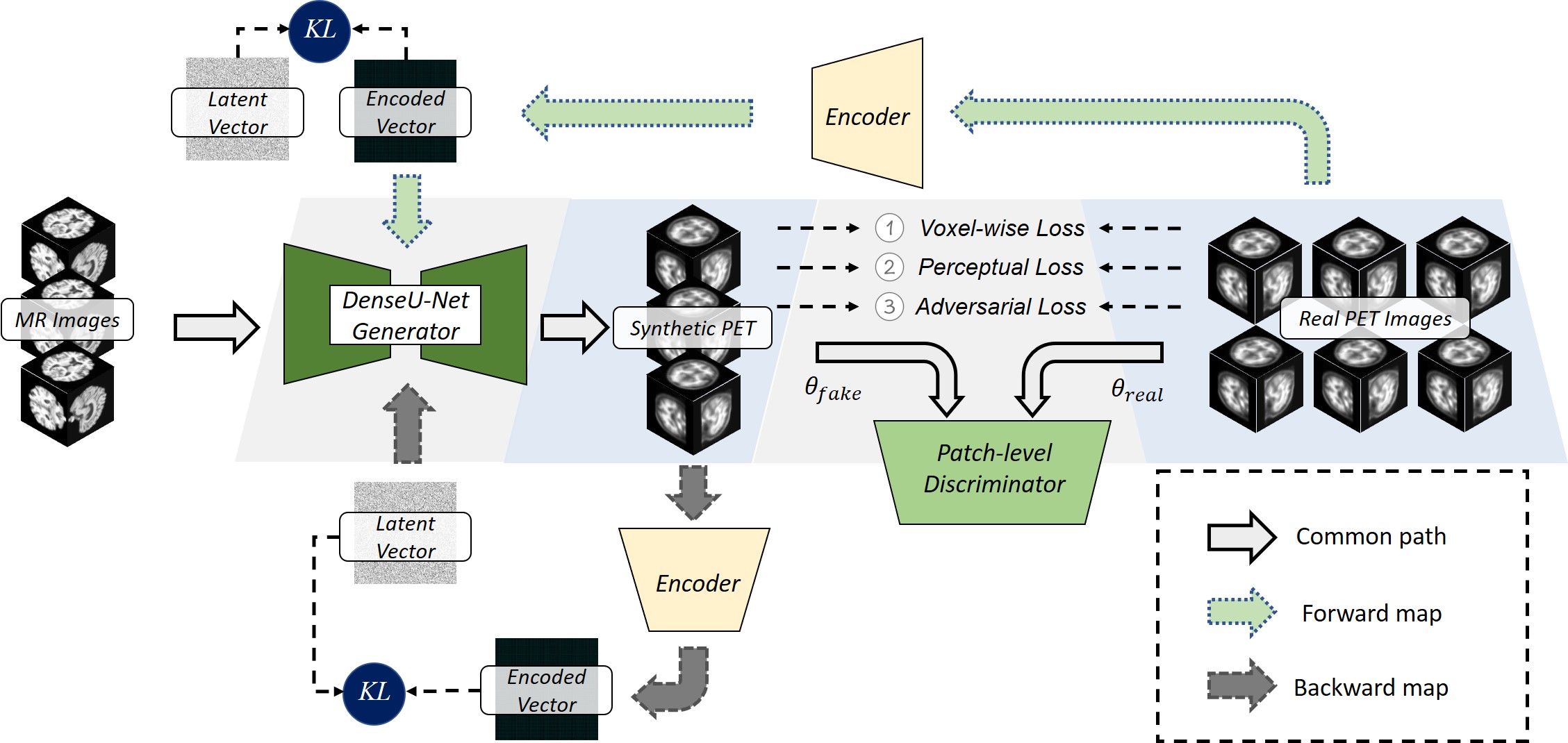}}
\caption{An overview of the proposed BMGAN for brain MR-to-PET synthesis.}
\label{fig1}
\end{figure*}

Assuming have a dataset containing the paired brain MR images $x \in \mathbb{R}_{MR}^{h \times w \times d}$  and PET images  $y \in \mathbb{R}_{PET}^{h \times w \times d}$ , our goal is learning a cross-modality mapping that can be formulated as $f: x \rightarrow y$ , where $f$ denotes the complex non-linear mapping between the brain MR and PET images. Nevertheless, there are diverse geometric structures among brain anatomies of different subjects, which introduce a huge challenge to the generative capability of cross-modality networks. To address this challenge, we creatively design a 3D generative adversarial network with the bidirectional mechanism. Our proposed BMGAN is illustrated in Fig.~\ref{fig1}, which consists of three components:1) the generator network, 2) the discriminator network, and 3) the encoder network. We firstly propose an advanced DenseU-Net generator that combines the architectures of U-Net~\cite{ref_41} and DenseNet~\cite{ref_42} to synthesize the targeted PET images from the corresponding brain MR images. It should be stressed that the 3D convolutional operations are adopted to optimally model the spatial structural information of PET images and eliminated the slices discontinuity problem caused by 2D networks. Then we utilize the adversarial learning strategy for the designed network, where an additional patch-level discriminator network is modeled. Last but not least, we introduce a bidirectional mapping mechanism by adding an encoder network to embed the semantic information of PET images into the latent space, which encourages the generator to preserve the diverse details of brain structures in synthetic PET images. Moreover, our generator is featured by incorporating the adversarial loss, the reconstruction loss, and the perceptual loss into the objective functions, with the goal of improving the visual quality of the synthetic target images. The details of the basic ideas, the architecture, and the extensive objective functions are described in the following Sections 2.1, 2.2 and 2.3.

\section{Basic ideas of BMGAN}
GANs~\cite{ref_13} are the most popular generative models that have two sub-networks: a generator $G$ and a discriminator $D$. In such cases, $G$ learns to capture the real data distribution and maps the latent vector to the synthetic images. And $D$ learns to estimate the classification probability of input samples that came from the real image rather than the synthetic images. During its adversarial training mechanism, both $G$ and $D$ are trained simultaneously, with the generator is trained to generate synthetic images that cannot be distinguished from the real images, while the discriminator aims to distinguish the synthetic and real images. Upon convergence, $G$ is capable of producing realistic counterfeit images that $D$ cannot recognize~\cite{ref_13}.

Mapping from a high-dimensional MR input to a high-dimensional PET output is challenging. In order to guarantee the variability of the generated results, the latent vector sampled from a known distribution (such as a standard normal distribution) is usually set as an input of the generator. In this regime, the biggest advantage of the proposed model is that it provides an invertible connection between the brain PET images and the latent vectors, which encourages the generator to synthesize the perceptually realistic PET images while preserving the diverse details of brain structures in different subjects. Specifically, the proposed method not only learn a forward mapping from the latent vector to the PET images like traditional GANs, but also learn a backward mapping that returns the PET images back to the latent space by training an encoder at the same time. This mechanism explicitly encourages a bidirectional consistency between the latent space and the PET images, so as to the semantic information of PET images are embedded into the high-dimensional latent space.

In the training phase of our BMGAN, the forward mapping starts with encoding the PET images into the latent space. To ensure the encoded latent vector have a similar distribution with the sampled latent vector, the encoded vector is trained to conform to the standard normal distribution by optimizing the KL divergence. The generator then tries to map the input MR images and the encoded vector to the synthetic PET images. In the backward mapping, the generator is first used to synthesize PET images from the MR images along with the sampled latent vector. Subsequently, the synthetic PET image is fed to the encoder to attempt to reconstruct the input latent vector. Combining the bidirectional mapping, the proposed method can enforce the generator network to utilize the latent code with the semantic information of PET images. Here as well, the role of discriminator is to try to distinguish the synthetic PET images from the real PET images. In the inference phase of our BMGAN, a deterministic generator uses the MR input images along with the sampled latent vectors to synthesize the corresponding PET images.

\subsection{Architectures}
\subsubsection{DenseU-Net Generator Network}
The generator architecture is paramount to the quality of synthetic images. The CNN or FCN architectures are used as the generator in many computer vision researches. Since the original MR images and the targeted PET images belong to the same subject, there are numerous low-level features shared between them. But using the plain CNN or FCN architectures may lead to the information loss of low-level structural features with the network depth increasing, which is harmful for the medical image synthesis. The emergence of U-Net effectively addresses this difficulty. It is comprised of the contracting subnetwork and expanding subnetwork to facilitates the image processing at multi-scales and has been widely applicable for segmenting medical images due to it can be properly trained using little data comparing to the original FCN. Since the researches of medical image synthesis address the pixel-level prediction issue like the image segmentation, it is a practicable attempt to adopt U-Net architectures in medical image synthesis tasks. In particularly, the contracting paths between them ensure the fusion of high-level feature and low-level information, which enhance the spatial and structural information of the synthetic images. Therefore, the U-Net architecture is very suitable as a generator in medical image synthesis tasks, and its effectiveness has been proved in many prior works~\cite{ref_15, ref_21, ref_24}.

But as the model becomes increasingly deeper, an awkward problem emerges: when the input or gradient information passes through many network layers, it will vanish and dilute. To tackle this problem, the short paths from earlier layers to later layers should be enforced. And one of the most effective models in this part is the DenseNet~\cite{ref_42}. The advantage of using DenseNet over other methods like ResNet~\cite{ref_35} has been well demonstrated in both accuracy and efficiency of many computer vision tasks.

Inspired by the dense connections, we propose a 3D DenseU-Net generator in this work. The main idea of the proposed DenseU-Net is to combine the advantages of both DenseNet and U-Net together. Let $x_{\ell}$ represent the output of $\ell$ th layer. The dense connection can be formally expressed as:
\begin{equation}x_{\ell}=H_{\ell}\left(\left[x_{\ell-1}, x_{\ell-2}, \ldots, x_{0}\right]\right)\label{eq}\end{equation}
where $H_{\ell}$ are the convolutional operations, and $\left[x_{\ell-1}, \cdots, x_{0}\right]$ represents the cascade concatenation.

The main difference the DenseU-Net from the naive U-Net is that the densely connected paths between layers are employed within each convolutional blocks of the encoder or decoder part. That means the dense blocks are formed of densely connecting the convolution layers with the same resolution in encoder and decoder part. Hence, the information flow and parameters efficiency of network can be improved for easily training the synthetic models, benefitting from the dense connection. At the same time, the long-range skip connection of U-Net improve the low-level structural feature preservation for better MR-to-PET synthesis.

\begin{figure*}[!t]
\centerline{\includegraphics[width=2\columnwidth]{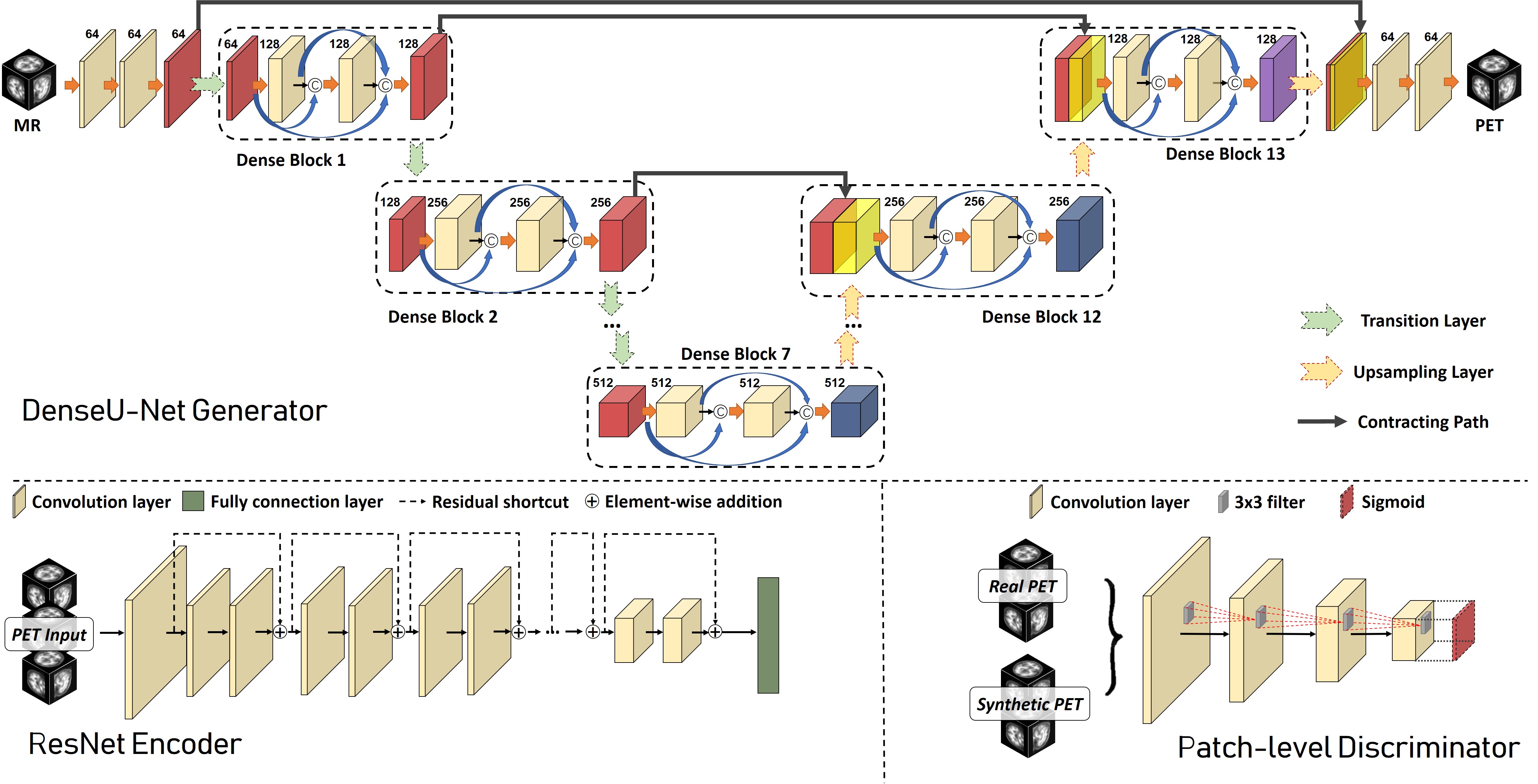}}
\caption{The illustration of our network architecture.}
\label{fig2}
\end{figure*}

The proposed DenseU-Net consists of 13 dense blocks, 7 transition layers and 7 upsampling layers, see Fig.~\ref{fig2}. Each dense block includes two directly connected convolutional layers. The filter kernel size of each convolutional layer is $3 \times 3 \times 3$. And the Leaky ReLU with the 0.2 negative slope is selected as the activation function. The number of the filters in each convolutional layer is marked in Fig.~\ref{fig2}. To halve the size of output feature maps, the transition layers consisted of a $1 \times 1 \times 1$ convolution layer followed by a max-pooling layer are used in the contracting subnetwork. And the compression rate of dense connection set as 1 in our experiments. In the expanding subnetwork, the upsampling layers with the factor of 2 are implemented by the transpose-convolutional layers. Using the contracting paths, the feature maps from the contracting subnetwork are concatenated with the feature maps of the expanding subnetwork, as denoted by the black straight arrows in Fig.~\ref{fig2}. For the normalization layers, we use the instance normalization rather than batch normalization to largely ease the optimization and benefit the generalization of deep networks. It has been proved to have a better performance in image synthetic tasks~\cite{ref_33, ref_34}. In the output layer, the normalization layer is removed and the Tanh function is adopted to obtain the synthetic PET images.
\subsubsection{Patch-level Discriminator Network}
For the discriminator, a typical CNN architecture including the convolution layer, rectified linear unit (ReLU) activation function and the max-pooling operation is utilized. As shown in Fig.~\ref{fig2}, the patch-level discriminator that classifies if each $32 \times 32 \times 32$ patch of the input image is real or fake is employed rather than the plain pixel-level structure. This structure not only has a few parameters, but also can accurately reconstruct the local brain structure of PET images. The filter size is $3 \times 3 \times 3$, and the numbers of the filters are 32, 64, 128 and 256 for the four convolutional layers, respectively. In the last convolutional layer, sigmoid activation function is used rather than ReLU function to represents the likelihood that the input is drawn from the real images.
\subsubsection{ResNet Encoder Network}
For the encoder, the ResNet architecture is deployed to instead of plain CNN to better encode the images. We follow the ResNet-34 network architecture with multiple Convolution-Normalization-Relu-Pooling components and several residual mapping to build the encoder network. The details of ResNet-34 can be found in~\cite{ref_35}.

\subsection{Objective functions}
Success in imaging synthesis tasks requires semantic reasoning. For MR-to-PET synthesis the synthetic PET output must be semantically similar to the corresponding MR input despite huge changes in appearance. In order to further improve the quality of generated PET images, an extensive objective function is used to optimize the network, including four types of terms: the adversarial loss, the KL-divergence constraint, the pixel-wise reconstruction loss, and the perceptual loss. The details will be introduced as follow.

The significance of the adversarial loss is to match the synthetic data distribution to the target data distribution. During the adversarial training process of GANs, the generator and discriminator are learned simultaneously. However, the adversarial objective of vanilla GAN cannot exactly reveal the inconsistency between the synthetic data distribution and the target data distribution. In such situation where the information between the original domain and target domain differs too much (such as MR-to-PET), the adversarial objective of vanilla GAN is hardly able to optimally converge, thus suffering from the mode collapse problem.

To deal with this challenge, the least-square adversarial objective function ~\cite{ref_30} is adopted to replace the objective of vanilla GAN in the proposed BMGAN, which turns out to be stable enough, thus remarkably addressing the convergence or mode collapse problem. Given an MR slice $x \sim P_{MR}(x)$ and its corresponding PET image $y \sim P_{PET}(y)$, the loss objective can be expressed as (2), (3):
\begin{equation}
\begin{split}
\mathcal{L}_{Ours}(D)&=\mathbb{E}_{x \sim P_{M R}(x), z \sim P_{z}(z)}\left[(D(G(x, z)))^{2}\right]\\
&+\mathbb{E}_{y \sim P_{PET}(y)}\left[(D(y)-1)^{2}\right]
\end{split}
\label{eq}
\end{equation}
\begin{equation}\mathcal{L}_{Ours}(G)=\mathbb{E}_{x \sim P_{M R}(x), z \sim P_{z}(z)}\left[(D(G(x, z))-1)^{2}\right]\label{eq}\end{equation}

To ensure the encoded vector have a similar distribution with the sampled latent vector, we enforce the KL-divergence constraint in encoder network, which means the difference between the encoded vector and the gaussian latent vector should be minimized, as shown below:
\begin{equation}\mathcal{L}_{Ours}(C)_{\text {For}}=\mathbb{E}_{y \sim P_{PET}(y)}\left[\emptyset_{KL}(\mathbb{E}(y) | \mathbb{N}(0,1))\right]\label{eq}\end{equation}
\begin{equation}
\mathcal{L}_{Ours}(C)_{Back}=\mathbb{E}_{x \sim P_{MR}(x), z \sim P_{z}(z)}\left[\emptyset_{KL}(\mathbb{E}(G(x, z))|\mathbb{N}(0,1))\right]
\label{eq}
\end{equation}
where E denotes expected value, $\emptyset$ denotes KL divergence and $z \sim P_{z}(z)$ represents sampling a latent vector from the Gaussian latent space.

Minimizing the adversarial objective function means that the generator G is making the discriminator D as confused as possible, in the sense that D cannot correctly discriminate the synthetic images and the real images. However, it is irresponsible for guaranteeing the structure of synthetic regions is consistent with that of real regions. For example, G always generate PET images that can confuse D but without being close to the real images. To address the problem, we introduce some extensive objective functions.

First, an $\mathcal{L}_{1}$ pixel-wise reconstruction loss is incorporated to impose an additional structural constraint on the generator. That means the generator is not only needed to fool the discriminator, but also needed to minimize the absolute pixel-wise intensity difference between synthetic PET images and real PET images. The $\mathcal{L}_{1}$ pixel-wise reconstruction loss is formalized as follows:
\begin{equation}\mathcal{L}_{1}(G)=\mathbb{E}_{x \sim P_{M R}(x), z \sim P_{z}(z), y \sim P_{P E T}(y)}\|y-G(x, z)\|_{1}\label{eq}\end{equation}

Unfortunately, despite the pixel-wise loss is able to capture the overall structure, it fails to capture perceptual differences between the synthetic and real images. For example, considering two identical MR images offset from each other by one pixel, they would be very different as measured by pixel-wise loss despite their perceptual similarity~\cite{ref_31}. Accordingly, a perceptual loss that depends on the dissimilarities between high-level feature representations is further introduced to generate PET images the higher quality, which is formalized as follows:
\begin{equation}
\begin{split}
\mathcal{L}_{\text {Perceptual}}(G)&=\mathbb{E}_{x \sim P_{MR}(x), z \sim P_{z}(\mathbf{z}), y \sim P_{P E T}(y)}\|V(y)\\
&-V(G(x, z))\|_{1}
\label{eq}
\end{split}
\end{equation}
where $V$ is a set of feature maps right before the second max-pooling operation of pre-trained VGG-16 [32].

The aggregate objective functions of generator G can be written as:
\begin{equation}
\begin{split}
\mathcal{L}_{Ours}(G)&=\mathbb{E}_{x \sim P_{MR}(x), z \sim P_{z}(z)}\left[(D(G(x, z))-1)^{2}\right]\\
&+\lambda_{1} \mathcal{L}_{1}(G)+\lambda_{2} \mathcal{L}_{\text {Perceptual}}(G)
\label{eq}
\end{split}
\end{equation}
where $\lambda_{1}$ and $\lambda_{2}$ are the hyper-parameters for controlling the relative importance of individual loss terms.

\section{Experiment}

\subsection{Experiment settings}
Alzheimer’s disease (AD) is a neurological and irreversible brain disease. The accurate diagnosis combining the different imaging modalities such as MR and PET plays a significant role in patient care, especially at the early stage. One of the impactive public neuroimaging datasets for the AD diagnosis is the Alzheimer’s Disease Neuroimaging Initiative (ADNI)~\cite{ref_38}, which includes MR, PET, and other imaging of different modalities.

The evaluated data of the proposed method was obtained from a subset of ADNI database, including 680 subjects. And each subject had images of two modalities (i.e. MR and PET). In the experiments, the proposed method is to learn the non-linear mapping from the MR images to the FDG-PET images. To avoid the influence of redundant information, the non-brain tissues between the MR images and PET images were removed. Moreover, the two modalities were spatially aligned to the same standardized template space to make them rigidly aligned with each other. To reduce the computational cost, we resized both the MR images and PET images to 128 × 128 × 128 voxels. To make full use of the available data, the 10-fold cross-validation was performed, where 7 folds were used as the training set, 1 fold for the validation set and remaining 2 fold was used as the test set. The proposed BMGAN method was implemented in Python, using the PyTorch deep learning framework. And all experiments were conducted out on an NVIDIA GeForce GTX 2080 Ti GPU.

In this work, we employ the following four metrics to evaluate the performance in the synthetic medical images, including the mean absolute error (MAE), the peak-signal-to-noise-ratio (PSNR), multi-scale structural similarity (MS-SSIM)~\cite{ref_39}, and Freshet Inception Distance (FID)~\cite{ref_40}.

To quantitatively evaluate the model performance in means of reconstruction we use MAE:
\begin{equation}MAE=\frac{1}{N} \sum_{i=1}^{N}\left|R_{PET}(i)-S_{PE T}(i)\right|\end{equation}
where $R_{PET}$ and $S_{PET}$ represents the real PET images and the synthetic PET images, respectively. And $i$ is the index of image pixels.

In addition, PSNR has been also widely used to measure the quality of synthetic images:
\begin{equation}P S N R=10 \log _{10} \frac{20^{2}}{MSE}\end{equation}
where $MSE$ is the mean-squared error between the real PET images and the synthetic PET images.

One of the most successful evaluation metrics about image similarity is MS-SSIM. It attempts to extract the multi-scale structural information from the image, where the higher MS-SSIM represents the better similarity between the real PET and synthetic PET.
Let $x$ and $y$ represents two image patches extracted from the same spatial location from the synthetic PET and real PET, respectively), and let $\mu_{\mathrm{x}}$, $\sigma_{x}$ and $\sigma_{xy}$ be the mean of $x$, the variance of $x$, and the covariance of $x$ and $y$, respectively. For pixel $i$, SSIM is defined as
\begin{equation}\begin{aligned}
\operatorname{SSIM}(i) &=\frac{2 \mu_{x} \mu_{y}+C_{1}}{\mu_{x}^{2}+\mu_{y}^{2}+C_{1}} \cdot \frac{2 \sigma_{x y}+C_{2}}{\sigma_{x}^{2}+\sigma_{y}^{2}+C_{2}} \\
&=l(i) \cdot cs(i)
\end{aligned}\end{equation}
where $C_{1}$ and $C_{2}$ are the small constants avoiding zero denominator errors.
However, the SSIM metric is a single-scale approach. MS-SSIM is proposed to measure the multi-scale structural similarity.
\begin{equation}\operatorname{MS-SSIM}(i)=l_{M}^{\alpha}(i) \cdot \prod_{j=1}^{M} cs_{j}^{\beta_{j}}(i)\end{equation}
where $l_{M}$ and $cs_{j}$ are the terms we defined in SSIM at scale $M$ and $j$, respectively. ${\alpha}$ and $\beta_{j}$ are the constants used to adjust the relative importance of different components, according to~\cite{ref_39}.

The Fréchet Inception Distance (FID) is a special GAN metric to measures the quality and diversity of the synthetic images using the pre-trained Inception network [Rethinking the inception architecture for computer vision], where the lower FID the better performance.
 \begin{equation}\operatorname{FID}=\left\|\mu_{r}-\mu_{g}\right\|+\operatorname{Tr}\left(\mathbf{C}_{r}+\mathbf{C}_{g}-2\left(\mathbf{C}_{r} \mathbf{C}_{g}\right)^{1 / 2}\right)\end{equation}
where $\mu_{r}$ and $\mu_{g}$ are the empirical means of real gaussian variables and synthetic gaussian variables, which are modeled by the pre-trained Inception networks. Likewise, $\mathbf{C}_{r}$ and $\mathbf{C}_{g}$ are the empirical covariance of real gaussian variables and synthetic gaussian variables, respectively.

\subsection{Comparison with 2D-based BMGAN architectures}
\begin{figure*}[htb]
\centerline{\includegraphics[width=1.5\columnwidth]{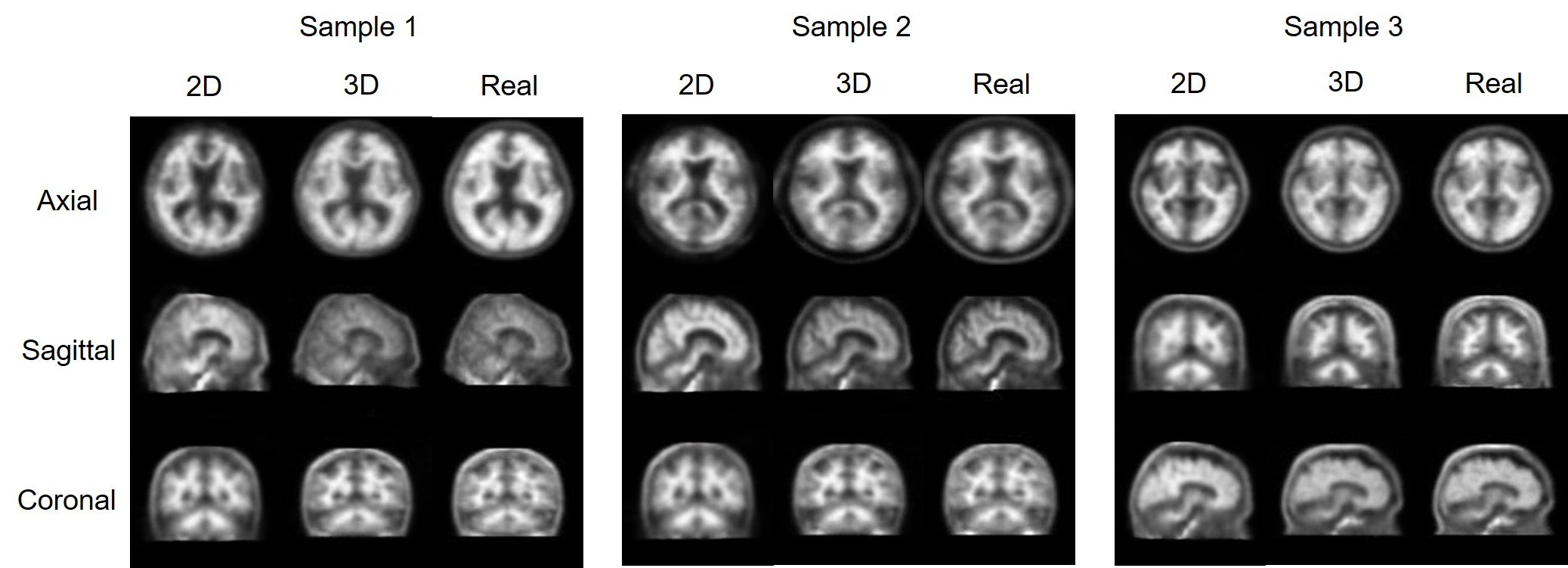}}
\caption{Qualitative comparison between synthetic results from 2D BMGAN and 3D BMGAN.}
\label{fig_e1_1}
\end{figure*}

To appraise the effectiveness of the 3D model, we compared it with the 2D variant of proposed model using the same network architectures in this section. The 2D BMGAN was trained with the slices of the axial, coronal, and sagittal planes. And it followed the same settings exploited in the proposed 3D model. It was noted that the 2D models takes the slices of one plane as input while 3D model uses the whole image as input. To qualitatively compare the results, the synthetic PET samples from the 2D BMGAN or the 3D BMGAN are shown in Fig.~\ref{fig_e1_1}.

\begin{figure}[h]
\centerline{\includegraphics[width=0.8\columnwidth]{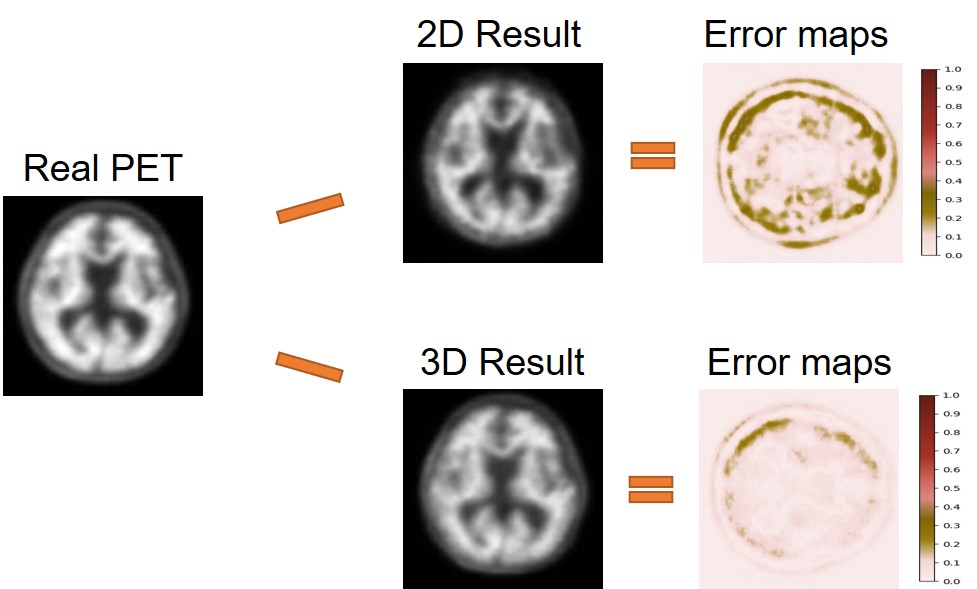}}
\caption{The difference maps between the synthetic PET and ground-truth PET from the axial plane of one subject.}
\label{fig_e1_2}
\end{figure}

For illustration, we show the synthetic results of three subject samples between the proposed 3D method and the 2D variant (see Fig.~\ref{fig_e1_1}). And the real PET images are also provided. As can be seen, compared to 2D variant, the proposed 3D BMGAN synthesized better visually pleasing results that have more similar appearance to the ground truth PET image in all three planes. To prove this more clearly, the difference maps between the synthetic images and the real PET are shown in Fig.~\ref{fig_e1_2}. The pixel value in the difference maps is the absolute intensities difference between the synthetic images and real PET images. The pseudo-color processing is employed to emphasize the difference. Looking at the error maps in Fig.~\ref{fig_e1_2}, the result of 2D variant has larger high-difference regions compared to the result of 3D proposed model. This may be because there is a problem of discontinuous estimation across slices in 2D synthetic model. By considering the 3D volumetric patches rather than slices by slices, the proposed 3D BMGAN could better synthesize PET images by taking volume structural information into account.

\begin{table}[htb]
\label{table}
    \renewcommand\arraystretch{2}
    \begin{center}
    \caption{Quantitative comparison between 2D BMGAN and 3D BMGAN}
    \normalsize
    \resizebox{\columnwidth}{!}{
    \begin{tabular}{ccccc}
    \toprule
        Model & MAE        & PSNR       & MS-SSIM   & FID         \\\midrule
        2D    & 31.76±8.22 & 26.26±1.33 & 0.77±0.12 & 48.68±10.47 \\
        3D    & 25.34±7.84 & 27.88±1.17 & 0.89±0.08 & 29.17±8.81 \\
    \bottomrule
    \end{tabular}}
    \end{center}
    \label{biaoge_1}
\end{table}

The averaged quantitative results achieved by the 2D and 3D models are presented in Table 1, in terms of four different evaluation metrics. We found that 3D BMGAN has higher PSNR and SSIM, lower MAE and FID than 2D variant, which means it outperform the 2D variant in both image quality and diversity. Both qualitative and quantitative results demonstrate the benefits of employing the 3D convolutional operations over the commonly used 2D operations.

\subsection{Contribution of the Adversarial training strategy}
To study the efficacy of the adversarial training mechanism in the proposed model, we specially conduct comparison experiments between the proposed 3D BMGAN and the simplified model that remove the discriminator network (i.e., just the 3D DenseU-Net generator and the encoder network).

\begin{figure*}[!t]
\centerline{\includegraphics[width=1.5\columnwidth]{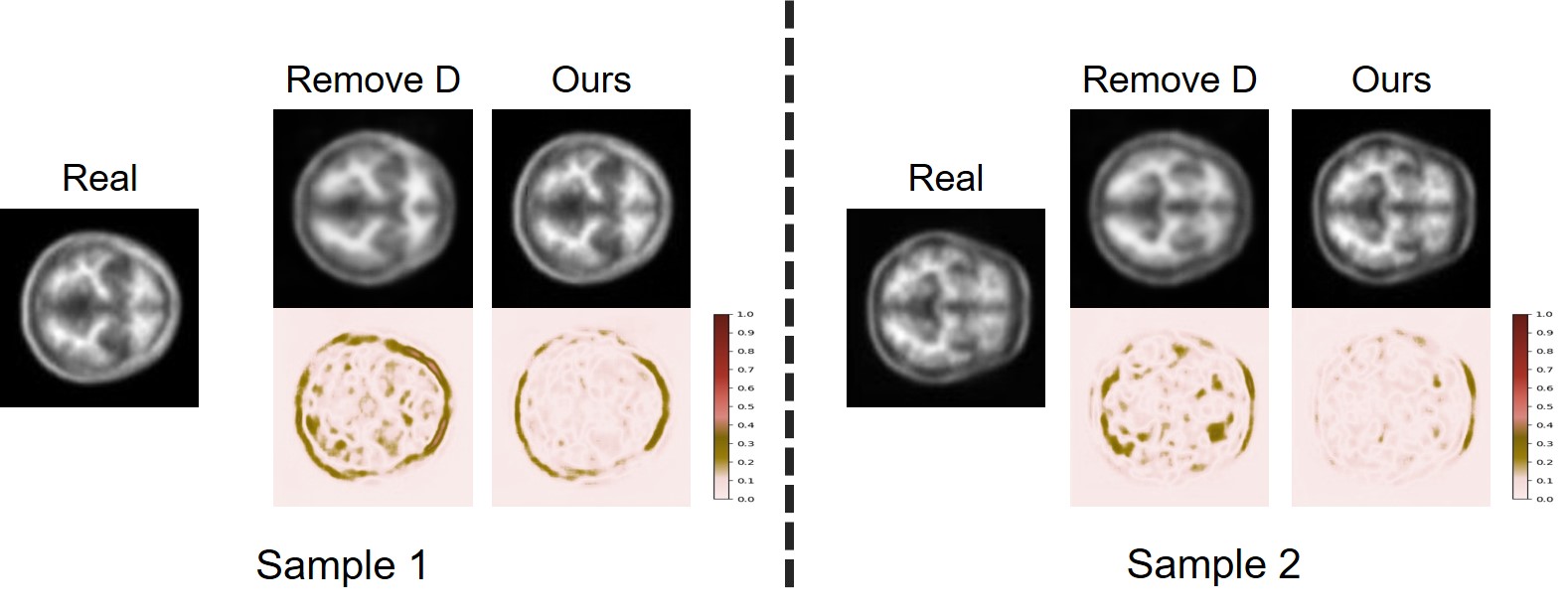}}
\caption{Qualitative comparison between the proposed BMGAN (Ours) and the simplified model that removed the discriminator network (Remove D).}
\label{fig_e2}
\end{figure*}

The axial slices of two synthetic 3D PET samples are visualized in Fig.~\ref{fig_e2}, where the synthetic images, real PET images and the difference maps are provided at once. We can clearly see that the synthetic images without the adversarial training strategy are easily over-smoothed and fuzzy compared to with the ground truth PET images. We speculate that this is because the simplified model without the adversarial training strategy synthesizes the PET images in a voxel-wise manner, which ignores the global structure and the context information. In this voxel-wise estimation manner, each voxel of the synthetic images is determined by the averaged value of potential overlapping regions. With the adversarial training strategy, the results are clearer and with better visual quality.

\begin{table}[htb]
\label{table}
    \renewcommand\arraystretch{2}
    \begin{center}
    \caption{Quantitative comparison between the proposed model (Ours) and the simplified model that removed the discriminator network (Remove D)}
    \normalsize
    \resizebox{\columnwidth}{!}{
    \begin{tabular}{ccccc}
    \toprule
        Model & MAE        & PSNR       & MS-SSIM   & FID         \\\midrule
        Remove D & 37.41±8.42 & 25.54±1.32 & 0.75±0.14 & 53.40±11.51 \\
        Ours     & 25.34±7.84 & 27.88±1.17 & 0.89±0.08 & 29.17±8.81 \\
    \bottomrule
    \end{tabular}}
    \end{center}
    \label{biaoge_2}
\end{table}

Table 2 show the quantitative results. We can see that adversarial training strategy brings obvious improvement in all metrics. For example, compared with the simplified model removed the discriminator network, the averaged PSNR of proposed method increases approximately 2.34. The experimental results demonstrated that the adversarial training mechanism of GAN is useful for improving the quality of synthetic medical images, indicating the essentials of this mechanism in our 3D BMGAN.

\subsection{Contribution of the DenseU-Net generator architecture}
\begin{figure*}[!t]
\centerline{\includegraphics[width=1.5\columnwidth]{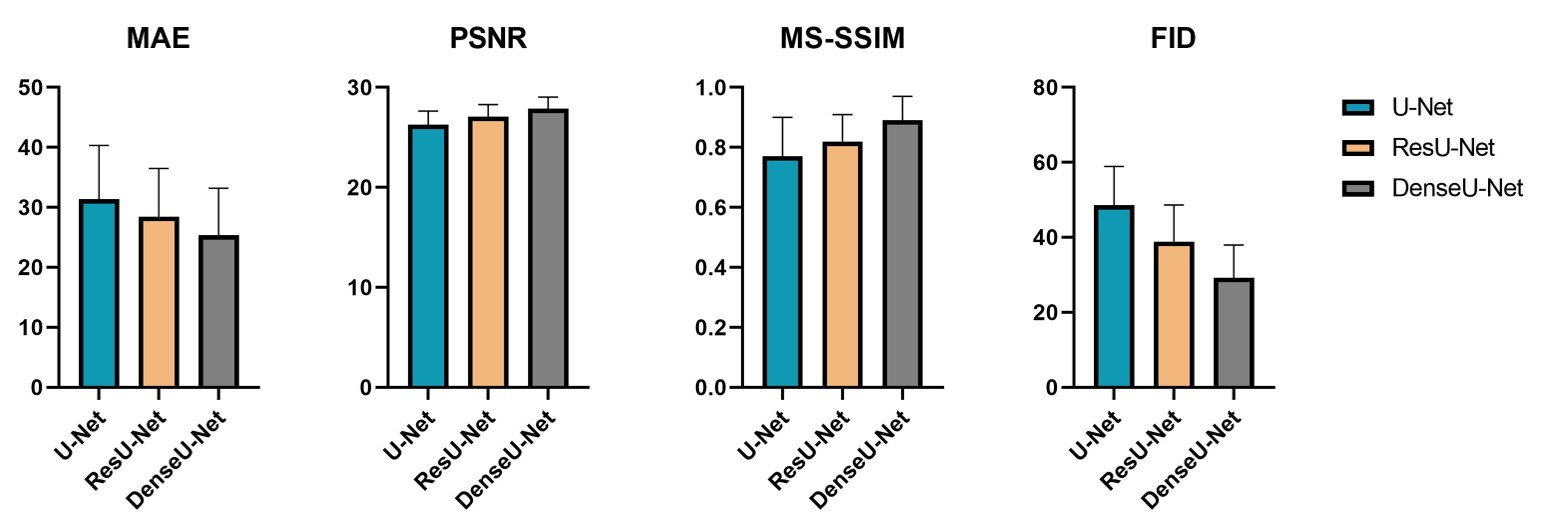}}
\caption{Quantitative comparison between three generator architectures, i.e. original U-Net, ResU-Net and DenseU-Net.}
\label{fig_e3}
\end{figure*}

As mentioned before, we design the DenseU-Net architecture, a variant of U-Net that combing the dense connections as the generator. To justify the effectiveness of the DenseU-Net, the performance of three generator architectures were compared in this section, i.e. original U-Net, ResU-Net and DenseU-Net. The detailed quantitative comparison in terms of MAE, PSNR, MS-SSIM and FID for the three architectures is given in Fig.~\ref{fig_e3}.

From Fig.~\ref{fig_e3}, we can see that, just using the original U-Net as the generator architecture obtains the lowest performance. The main reason is that the learnable parameters update difficultly due to the vanishing gradient issues. What’s more, the model is easier to slide into the local optimum. Therefore, just using the U-Net architecture including skip connection leads to unsatisfactory results. Compared to using the original U-Net, employ the ResU-Net including residual paths achieves a better quantitative result. This is due to the fact that the synthetic model might be difficult to converge to the optimal value during the training phase because of the vanishing gradients or the cost of GPU memory. In such cases, the residual learning can boost the gradient flow across different layers and make the network much easier to converge to the optimal parameters, which help synthesize the PET images from the MR images. Nevertheless, it can be seen that using the DenseU-Net architecture as the generator achieves the best performance in all metrics. The main reason triggering this improvement is that the feature reuse of dense connection is more efficient in the parameter and ensures the maximum information flow. It can be seen as an extension of residual learning. The inconsistency of the feature distributions between the MR and PET images could be mitigated by the dense block and skip connection. The results suggesting that the DenseU-Net generator architecture can further improve the image quality of the synthetic PET images.

\subsection{Contribution of the extensive objective functions}
As mentioned in the Method Section, the extensive objective function is employed to optimize the network, including the adversarial loss, the KL-divergence constraint, the pixel-wise reconstruction loss ($\mathcal{L}_{1}$ loss) and the perceptual loss. To study the contribution of the extensive objective functions for MR-to-PET synthesis, we use 1) adversarial loss and KL-divergence constraint, 2) adversarial loss, KL-divergence constraint, and $\mathcal{L}_{1}$ loss, 3) adversarial loss, KL-divergence constraint, $\mathcal{L}_{1}$ loss, and perceptual loss (ours) for training the proposed model, respectively. Results are shown in  Table 3. Note that we observed that there is a mode collapsed problem when using the adversarial loss of vanilla GAN to train the model. Therefore, the adversarial loss of LSGAN was used in the proposed model to slightly address the problem. For more information about the comparison of adversarial loss between LSGAN and vanilla GAN, please refer to the~\cite{ref_30}.

\begin{table}[htb]
\label{table}
    \renewcommand\arraystretch{2}
    \begin{center}
    \normalsize
    \caption{Quantitative comparison between different loss function}
    \resizebox{\columnwidth}{!}{
    \begin{tabular}{ccccc}
    \toprule
        Loss function    & MAE        & PSNR       & MS-SSIM   & FID        \\\midrule
        Adversarial+KL      & 38.57±9.13 & 24.57±1.34 & 0.67±0.14 & 69.55±18.12 \\
        Adversarial+KL+$\mathcal{L}_{1}$   & 27.81±7.99 & 27.26±1.19 & 0.84±0.10 & 36.73±9.27  \\
        Ours             & 25.34±7.84 & 27.88±1.17 & 0.89±0.08 & 29.17±8.81  \\
    \bottomrule
    \end{tabular}}
    \end{center}
    \label{biaoge_3}
\end{table}

In the trained model without the reconstruction loss and perceptual loss, the results seem tolerable but the quality of synthetic images is not very good. The main problems are that the adversarial loss and KL-divergence constraint focus on the global information and fails to pull the distribution between the real PET and synthetic PET closely enough in the pixel level.

The first kind of extension losses is the pixel-wise reconstruction loss ($\mathcal{L}_{1}$ loss) added to the adversarial loss and KL-divergence constraint. From the Table 3, adding the pixel-wise loss significantly improve the quality of synthetic images in all four metrics. It is helpful to capture the overall structure in relation to the PET context. Then, the second kind of extension losses is the perceptual loss. Compared to the previous result, adding the perceptual loss brings a modest improvement. And the improvement on MS-SSIM is relatively significant. This is because the perceptual losses help the model to keep more detailed feature representations of the synthetic PET images.

Above all, with the extensive objective functions including the adversarial loss, the KL-divergence constraint, the pixel-wise reconstruction loss and the perceptual loss, our model is able to performs stable and robust MR-to-PET synthesis on the current applications.

\subsection{Comparison with the competitive synthetic methods}
In this section, we compare our method with the existing medical imaging cross-modality synthesis models to prove its effectiveness. The original FCN and GAN was used as the baseline models. A typical 3D FCN is employed to perform this task, while the pooling operation is not used due to it potentially leads to the loss of information and resolution. Moreover, the compared GAN consists of the FCN generator and the classic discriminator. To further enhance the reliability, some comparisons with the models which have done specifically for MR-to-PET synthesis were made, such U-Net~\cite{ref_36} and Cycle-consistent GAN (CycleGAN)~\cite{ref_37}. Finally, we compared the synthetic results of proposed model with that of pGAN~\cite{ref_16}, which was the state-of-the-art model in cross-modality synthesis models. It was noted that we re-implemented these models according to their original papers, due to their codes were unavailable. The topology of models was retained, while some necessary changes were made to adapt to current task.

\begin{figure*}[!t]
\centerline{\includegraphics[width=1.5\columnwidth]{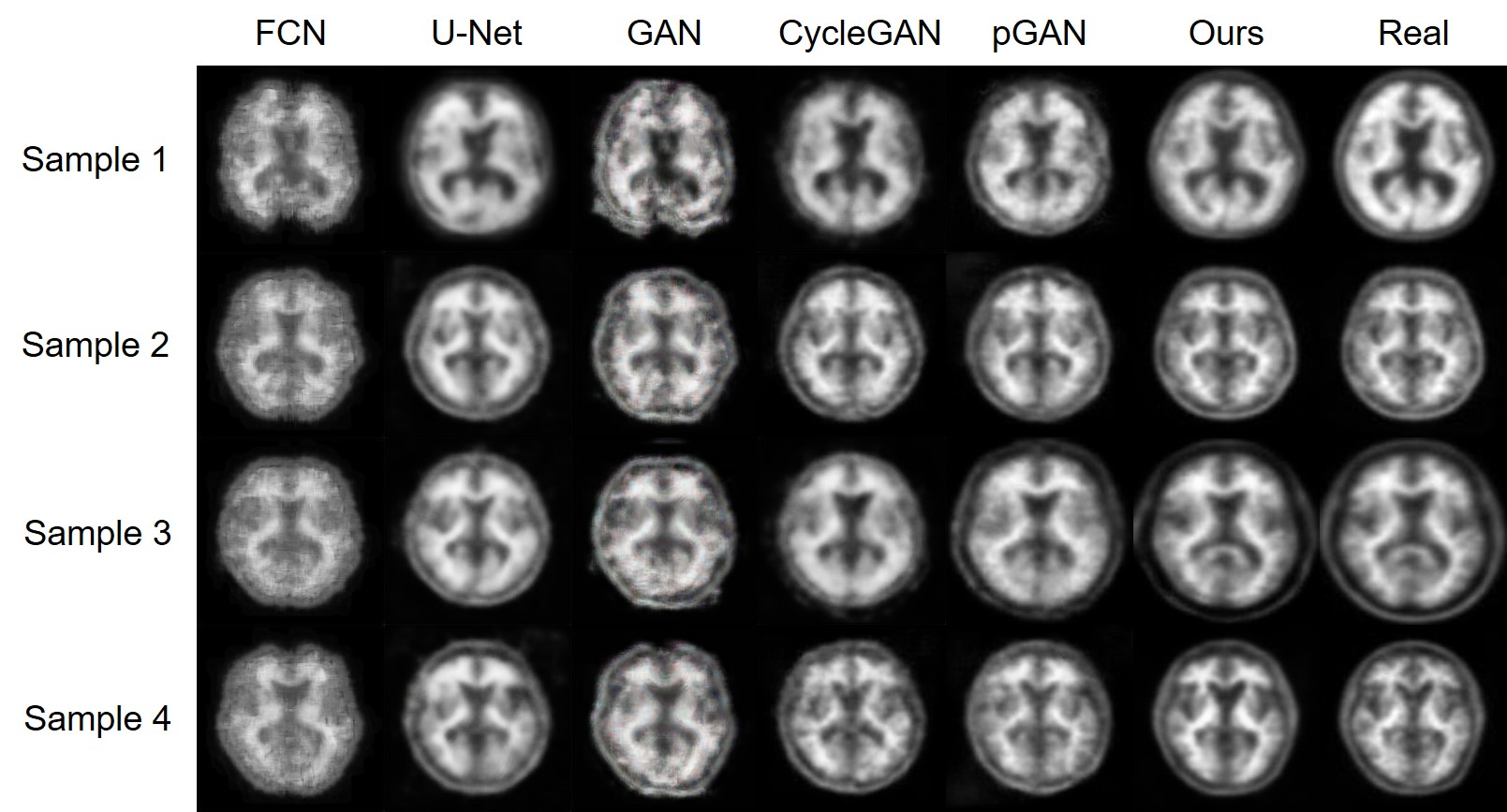}}
\caption{Qualitative comparison of original MR images, synthetic PET from the 3D FCN model, synthetic PET from the 3D GAN model, synthetic PET from the 3D cGAN model, and synthetic PET from the 3D CycleGAN model, as well as the real PET images.}
\label{fig_e5_1}
\end{figure*}

\begin{figure*}[!t]
\centerline{\includegraphics[width=1.5\columnwidth]{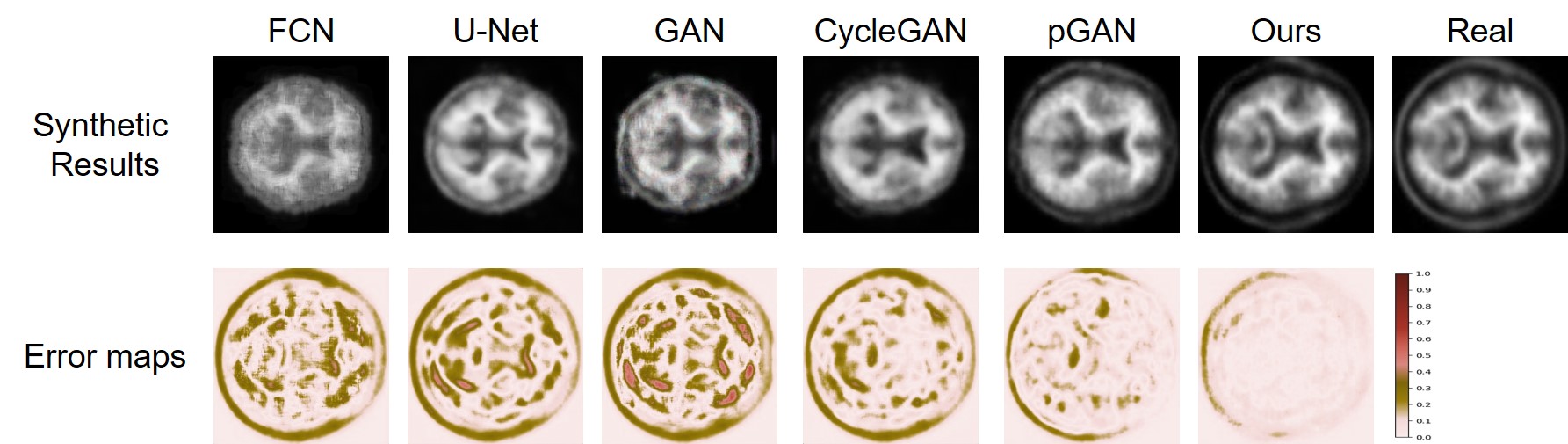}}
\caption{The difference maps between the synthetic PET and ground-truth PET in axial plane.}
\label{fig_e5_2}
\end{figure*}

 Fig.~\ref{fig_e5_1} is the illustrative figure showing the synthetic images from each model in axial planes. In  Fig.~\ref{fig_e5_1}, we show both the synthetic results and real PET images from four different samples, which have diverse brain structure. To clearly show the improvement of synthetic PET images, the difference maps of an axial image from each model were calculated (see Fig.~\ref{fig_e5_2}).

 As can be seen in  Fig.~\ref{fig_e5_1} and  Fig.~\ref{fig_e5_2}, there is a significant appearance difference between the output of the FCN and the ground truth images. This may be because the network structure restricts its generative capability under the current limited training dataset. Meanwhile, the synthetic results of U-Net roughly restore the skeleton of real PET images while with blurry estimations in structural details. We speculate that the advantage of U-Net compared to FCN is that it can be properly trained when the training samples is relatively limited, which is quite common in medical image analysis tasks. And the U-Net network architecture with the skip connection may be more suitable for the medical image analysis tasks because of the better preservation of context information. To proceed, compared with the FCN without adversarial training, the synthetic results of GAN using the FCN as the generator are closer to the real PET images, which further proves the advantages of adversarial training strategy. However, it should be noted that the quality of its synthetic images is still very poor, with many artifacts.

As the state-of-the-art synthetic models, the synthetic results of CycleGAN and pGAN obviously achieved better visual quality than the above three models. But comparing to the real PET images, we can observe that their synthetic results are coarser within the region close to the boundary, while in the results of our proposed BMGAN, these regions are sharper and clearer. What’s more, the synthetic images by our method appreciably improve the detail and diversity, which are closest to the real images between the multiple subjects with different brain structure. In contrast, the synthetic images of CycleGAN or pGAN have worse appearances in the case that the ground-truth images have large structural changes. And some wrong results may be produced. The potential reason is that the structural and pixel distribution between the brain PET from different subjects may have a huge difference. In the current limited training dataset, CycleGAN or pGAN tend to synthesize the plausible but similar results, resulting in poor diversity of generated images.

The superiority of the proposed BMGAN in this scenario are clearly shown. In the proposed BMGAN, the latent space are embedded into some priori semantic information about the PET images. In this way, even if the structures of ground-truth images are diverse, the latent vectors with semantic prior knowledge can provide more auxiliary information for the GAN model to help it synthesize realistic and diverse PET images. From the figure, we can conclude that the proposed BMGAN has satisfying performance on synthesizing PET images from MR images.

\begin{figure*}[!t]
\centerline{\includegraphics[width=1.5\columnwidth]{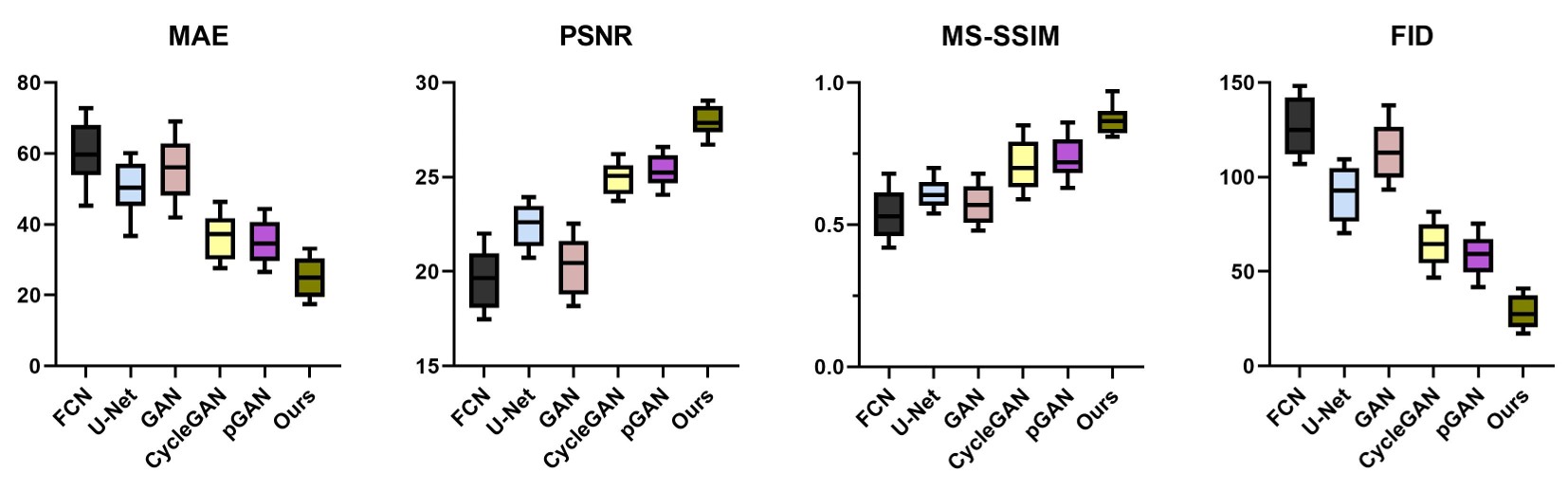}}
\caption{ Quantitative comparison between the existing synthetic methods and the proposed method.}
\label{fig_e5_3}
\end{figure*}

The quantitative comparison in terms of MAE, PSNR, MS-SSIM and FID metrics also provided in Fig.~\ref{fig_e5_3}. As observed, our proposed method achieves the best performance, which considerably outperforms other competitive methods. The lowest MAE and the highest PSNR represent the synthetic results by our method have the best quality, while the highest MS-SSIM and the lowest FID represent the synthetic results by our method have the best diversity. Both qualitative and quantitative results suggest that, compared with the competitive synthetic models, the proposed method can generate plausible and diverse PET images with better synthetic performance.

\begin{figure}[H]
\centerline{\includegraphics[width=0.5\columnwidth]{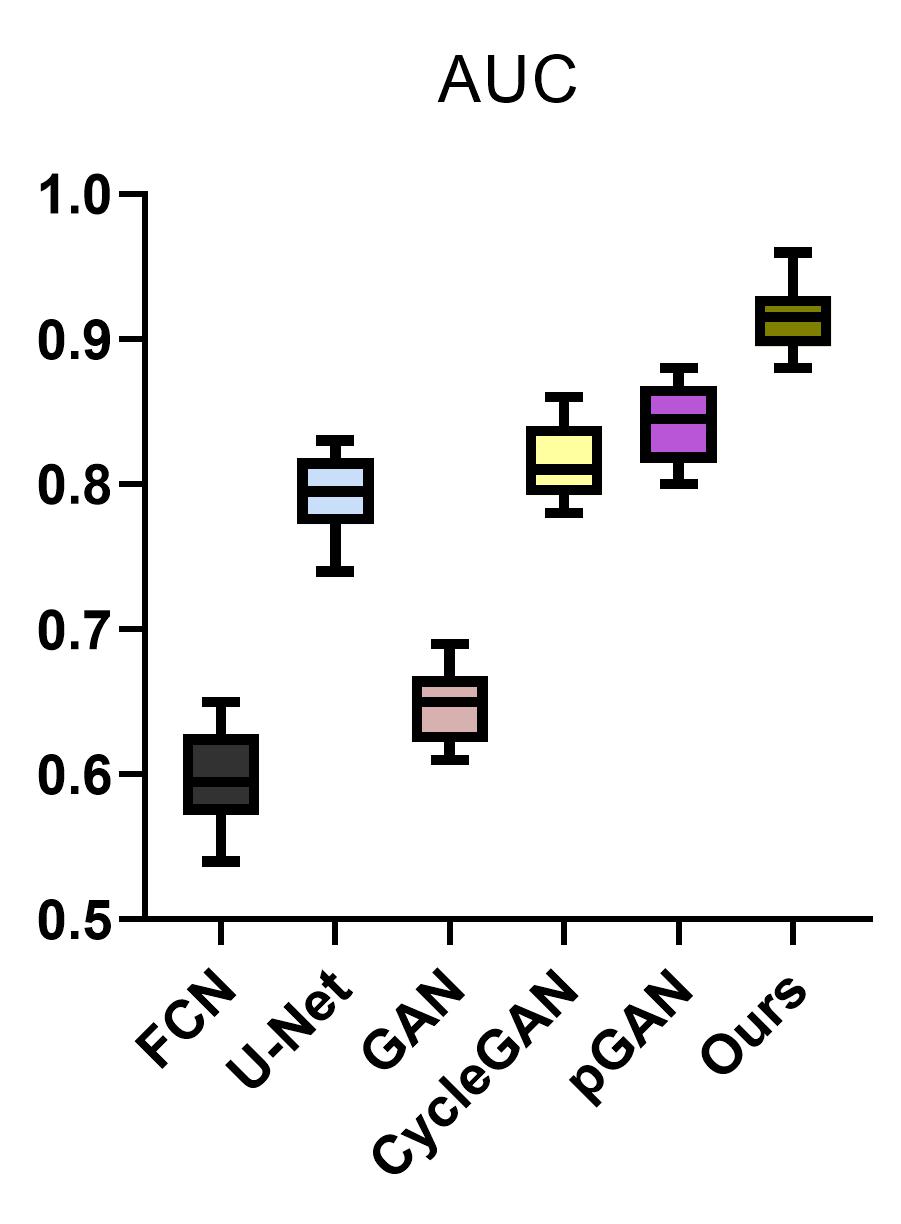}}
\caption{ Classification AUC of the synthetic PET images.}
\label{fig_e6}
\end{figure}

\subsection{Classification experiment on the synthetic images}
Nevertheless, the used evaluation metrics for synthetic images were not directly relevant to the diagnostic quality. To study the effectiveness of synthetic PET images in disease diagnosis quantitatively, we further calculated the classification accuracy of the synthetic PET images on a binary classification task (i.e. AD vs. Normal). VGG-16 network~\cite{ref_32}, a classic imaging classification networks in computer vision, is employed as the gold-standard classifier in the current study. For a fair comparison, we first pre-trained the VGG-16 using the real PET images. Then, the synthetic PET images from different methods were fed to the pre-trained VGG-16 to make a classification, respectively. The area under the ROC curve (AUC) was used as the evaluation metric to solve the class-imbalanced problem between different classes. And the results of classification experiment are shown in Fig.~\ref{fig_e6}. It can be observed that the AUC of the proposed BMGAN outperformed the other compared methods. According to the above experimental results, it can be concluded that the proposed BMGAN is capable of generating a substantial amount of realistic and diverse PET images with the detailed attributes of brains. And it also proved that the proposed BMGAN can be used as an effective data augmentation method.

\section{Conclusion}
In this work, we proposed a novel 3D BMGAN for synthesizing the brain PET images from the brain MR images. As an attempt to bridge the gap between the network generative capability and real medical images, the bidirectional mapping mechanism was introduced to encourage the generator to synthesize the perceptually realistic PET images while preserving the diverse details of brain structures in different subjects. Also, the 3D DenseU-Net generator architecture and the extensive objective functions were used to further improve the visual quality of synthetic PET images. The performance of our proposed method is evaluated in a subset of ADNI database. The experiment demonstrates that the diverse details of brain structures in different subjects can be faithfully preserved by our proposed method. Moreover, both qualitatively and quantitatively experimental results suggest that our proposed method can effectively synthesize PET images from MR images, outperforming the baseline deep learning methods and the existing state-of-the-art cross-modality methods. Please note that, the proposed method can be used in wider medical imaging synthesis applications where a cross-modality mapping from source domain to target domain is needed. In the future, we will further investigate the potential of incorporating the multi-modality information for better synthesis.


\begin{thebibliography}{99}

\bibitem{ref_1}
H.~Brody, ``Medical imaging,'' \emph{Nature}, vol. 502, no. 7473, pp. S81--S81, 2013.

\bibitem{ref_2}
V.~D. Calhoun and J.~Sui, ``Multimodal fusion of brain imaging data: a key to
  finding the missing link (s) in complex mental illness,'' \emph{Biological
  psychiatry: cognitive neuroscience and neuroimaging}, vol.~1, no.~3, pp.
  230--244, 2016.

\bibitem{ref_3}
M.~Liu, Y.~Gao, P.-T. Yap, and D.~Shen, ``Multi-hypergraph learning for
  incomplete multimodality data,'' \emph{IEEE journal of biomedical and health
  informatics}, vol.~22, no.~4, pp. 1197--1208, 2017.

\bibitem{ref_4}
H.-C. Shin, N.~A. Tenenholtz, J.~K. Rogers, C.~G. Schwarz, M.~L. Senjem, J.~L.
  Gunter, K.~P. Andriole, and M.~Michalski, ``Medical image synthesis for data
  augmentation and anonymization using generative adversarial networks,'' in
  \emph{International workshop on simulation and synthesis in medical
  imaging}.\hskip 1em plus 0.5em minus 0.4em\relax Springer, 2018, pp. 1--11.

\bibitem{ref_5}
N.~Burgos, M.~J. Cardoso, K.~Thielemans, M.~Modat, S.~Pedemonte, J.~Dickson,
  A.~Barnes, R.~M. Ahmed, C.~J. Mahoney, J.~M. Schott \emph{et~al.},
  ``Attenuation correction synthesis for hybrid pet-mr scanners: Application to
  brain studies,'' \emph{IEEE Transactions on Medical Imaging}, vol.~33,
  no.~12, pp. 2332--2341, 2014.

\bibitem{ref_6}
J.~Lee, A.~Carass, A.~Jog, C.~Zhao, and J.~L. Prince, ``Multi-atlas-based ct
  synthesis from conventional mri with patch-based refinement for mri-based
  radiotherapy planning,'' in \emph{Medical Imaging 2017: Image Processing},
  vol. 10133.\hskip 1em plus 0.5em minus 0.4em\relax International Society for
  Optics and Photonics, 2017, p. 101331I.

\bibitem{ref_7}
C.~Catana, A.~van~der Kouwe, T.~Benner, C.~J. Michel, M.~Hamm, M.~Fenchel,
  B.~Fischl, B.~Rosen, M.~Schmand, and A.~G. Sorensen, ``Toward implementing an
  mri-based pet attenuation-correction method for neurologic studies on the
  mr-pet brain prototype,'' \emph{Journal of Nuclear Medicine}, vol.~51, no.~9,
  pp. 1431--1438, 2010.

\bibitem{ref_8}
A.~Jog, A.~Carass, S.~Roy, D.~L. Pham, and J.~L. Prince, ``Random forest
  regression for magnetic resonance image synthesis,'' \emph{Medical image
  analysis}, vol.~35, pp. 475--488, 2017.

\bibitem{ref_9}
A.~Jog, A.~Carass, and J.~L. Prince, ``Improving magnetic resonance resolution
  with supervised learning,'' in \emph{2014 IEEE 11th International Symposium
  on Biomedical Imaging (ISBI)}.\hskip 1em plus 0.5em minus 0.4em\relax IEEE,
  2014, pp. 987--990.

\bibitem{ref_10}
T.~Huynh, Y.~Gao, J.~Kang, L.~Wang, P.~Zhang, J.~Lian, and D.~Shen,
  ``Estimating ct image from mri data using structured random forest and
  auto-context model,'' \emph{IEEE transactions on medical imaging}, vol.~35,
  no.~1, pp. 174--183, 2015.

\bibitem{ref_11}
R.~Li, W.~Zhang, H.-I. Suk, L.~Wang, J.~Li, D.~Shen, and S.~Ji, ``Deep learning
  based imaging data completion for improved brain disease diagnosis,'' in
  \emph{International Conference on Medical Image Computing and
  Computer-Assisted Intervention}.\hskip 1em plus 0.5em minus 0.4em\relax
  Springer, 2014, pp. 305--312.

\bibitem{ref_12}
L.~Xiang, Y.~Qiao, D.~Nie, L.~An, W.~Lin, Q.~Wang, and D.~Shen, ``Deep
  auto-context convolutional neural networks for standard-dose pet image
  estimation from low-dose pet/mri,'' \emph{Neurocomputing}, vol. 267, pp.
  406--416, 2017.

\bibitem{ref_13}
I.~Goodfellow, J.~Pouget-Abadie, M.~Mirza, B.~Xu, D.~Warde-Farley, S.~Ozair,
  A.~Courville, and Y.~Bengio, ``Generative adversarial nets,'' in
  \emph{Advances in neural information processing systems}, 2014, pp.
  2672--2680.

\bibitem{ref_14}
A.~Ben-Cohen, E.~Klang, S.~P. Raskin, S.~Soffer, S.~Ben-Haim, E.~Konen, M.~M.
  Amitai, and H.~Greenspan, ``Cross-modality synthesis from ct to pet using fcn
  and gan networks for improved automated lesion detection,'' \emph{Engineering
  Applications of Artificial Intelligence}, vol.~78, pp. 186--194, 2019.

\bibitem{ref_15}
L.~Xiang, Y.~Li, W.~Lin, Q.~Wang, and D.~Shen, ``Unpaired deep cross-modality
  synthesis with fast training,'' in \emph{Deep Learning in Medical Image
  Analysis and Multimodal Learning for Clinical Decision Support}.\hskip 1em
  plus 0.5em minus 0.4em\relax Springer, 2018, pp. 155--164.

\bibitem{ref_16}
S.~U. Dar, M.~Yurt, L.~Karacan, A.~Erdem, E.~Erdem, and T.~{\c{C}}ukur, ``Image
  synthesis in multi-contrast mri with conditional generative adversarial
  networks,'' \emph{IEEE transactions on medical imaging}, vol.~38, no.~10, pp.
  2375--2388, 2019.

\bibitem{ref_17}
P.~Isola, J.-Y. Zhu, T.~Zhou, and A.~A. Efros, ``Image-to-image translation
  with conditional adversarial networks,'' in \emph{Proceedings of the IEEE
  conference on computer vision and pattern recognition}, 2017, pp. 1125--1134.

\bibitem{ref_18}
D.~Pathak, P.~Krahenbuhl, J.~Donahue, T.~Darrell, and A.~A. Efros, ``Context
  encoders: Feature learning by inpainting,'' in \emph{Proceedings of the IEEE
  conference on computer vision and pattern recognition}, 2016, pp. 2536--2544.

\bibitem{ref_19}
P.~Sangkloy, J.~Lu, C.~Fang, F.~Yu, and J.~Hays, ``Scribbler: Controlling deep
  image synthesis with sketch and color,'' in \emph{Proceedings of the IEEE
  Conference on Computer Vision and Pattern Recognition}, 2017, pp. 5400--5409.

\bibitem{ref_20}
W.~Xian, P.~Sangkloy, V.~Agrawal, A.~Raj, J.~Lu, C.~Fang, F.~Yu, and J.~Hays,
  ``Texturegan: Controlling deep image synthesis with texture patches,'' in
  \emph{Proceedings of the IEEE Conference on Computer Vision and Pattern
  Recognition}, 2018, pp. 8456--8465.

\bibitem{ref_21}
Y.~Wang, B.~Yu, L.~Wang, C.~Zu, D.~S. Lalush, W.~Lin, X.~Wu, J.~Zhou, D.~Shen,
  and L.~Zhou, ``3d conditional generative adversarial networks for
  high-quality pet image estimation at low dose,'' \emph{NeuroImage}, vol. 174,
  pp. 550--562, 2018.

\bibitem{ref_22}
D.~Nie, X.~Cao, Y.~Gao, L.~Wang, and D.~Shen, ``Estimating ct image from mri
  data using 3d fully convolutional networks,'' in \emph{Deep Learning and Data
  Labeling for Medical Applications}.\hskip 1em plus 0.5em minus 0.4em\relax
  Springer, 2016, pp. 170--178.

\bibitem{ref_23}
D.~Nie, R.~Trullo, J.~Lian, L.~Wang, C.~Petitjean, S.~Ruan, Q.~Wang, and
  D.~Shen, ``Medical image synthesis with deep convolutional adversarial
  networks,'' \emph{IEEE Transactions on Biomedical Engineering}, vol.~65,
  no.~12, pp. 2720--2730, 2018.

\bibitem{ref_24}
Y.~Pan, M.~Liu, C.~Lian, T.~Zhou, Y.~Xia, and D.~Shen, ``Synthesizing missing
  pet from mri with cycle-consistent generative adversarial networks for
  alzheimer¡¯s disease diagnosis,'' in \emph{International Conference on
  Medical Image Computing and Computer-Assisted Intervention}.\hskip 1em plus
  0.5em minus 0.4em\relax Springer, 2018, pp. 455--463.

\bibitem{ref_25}
J.-Y. Zhu, R.~Zhang, D.~Pathak, T.~Darrell, A.~A. Efros, O.~Wang, and
  E.~Shechtman, ``Toward multimodal image-to-image translation,'' in
  \emph{Advances in neural information processing systems}, 2017, pp. 465--476.

\bibitem{ref_26}
S.~Hu, Y.~Shen, S.~Wang, and B.~Lei, ``Brain mr to pet synthesis via
  bidirectional generative adversarial network,'' in \emph{International
  Conference on Medical Image Computing and Computer-Assisted
  Intervention}.\hskip 1em plus 0.5em minus 0.4em\relax Springer, 2020.

\bibitem{ref_27}
A.~Krizhevsky, I.~Sutskever, and G.~E. Hinton, ``Imagenet classification with
  deep convolutional neural networks,'' in \emph{Advances in neural information
  processing systems}, 2012, pp. 1097--1105.

\bibitem{ref_28}
Y.~Li, S.~Tang, M.~Lin, Y.~Zhang, J.~Li, and S.~Yan, ``Implicit negative
  sub-categorization and sink diversion for object detection,'' \emph{IEEE
  Transactions on Image Processing}, vol.~27, no.~4, pp. 1561--1574, 2017.

\bibitem{ref_29}
L.~Li, S.~Tang, Y.~Zhang, L.~Deng, and Q.~Tian, ``Gla: Global--local attention
  for image description,'' \emph{IEEE Transactions on Multimedia}, vol.~20,
  no.~3, pp. 726--737, 2017.

\bibitem{ref_30}
X.~Mao, Q.~Li, H.~Xie, R.~Y. Lau, Z.~Wang, and S.~Paul~Smolley, ``Least squares
  generative adversarial networks,'' in \emph{Proceedings of the IEEE
  International Conference on Computer Vision}, 2017, pp. 2794--2802.

\bibitem{ref_31}
J.~Johnson, A.~Alahi, and L.~Fei-Fei, ``Perceptual losses for real-time style
  transfer and super-resolution,'' in \emph{European conference on computer
  vision}.\hskip 1em plus 0.5em minus 0.4em\relax Springer, 2016, pp. 694--711.

\bibitem{ref_32}
K.~Simonyan and A.~Zisserman, ``Very deep convolutional networks for
  large-scale image recognition,'' \emph{arXiv preprint arXiv:1409.1556}, 2014.

\bibitem{ref_33}
S.~Hu, J.~Yuan, and S.~Wang, ``Cross-modality synthesis from mri to pet using
  adversarial u-net with different normalization,'' in \emph{2019 International
  Conference on Medical Imaging Physics and Engineering (ICMIPE)}.\hskip 1em
  plus 0.5em minus 0.4em\relax IEEE, 2019, pp. 1--5.

\bibitem{ref_34}
J.-Y. Zhu, T.~Park, P.~Isola, and A.~A. Efros, ``Unpaired image-to-image
  translation using cycle-consistent adversarial networks,'' in
  \emph{Proceedings of the IEEE international conference on computer vision},
  2017, pp. 2223--2232.

\bibitem{ref_35}
K.~He, X.~Zhang, S.~Ren, and J.~Sun, ``Deep residual learning for image
  recognition,'' in \emph{Proceedings of the IEEE conference on computer vision
  and pattern recognition}, 2016, pp. 770--778.

\bibitem{ref_36}
K.~D. Spuhler, J.~Gardus, Y.~Gao, C.~DeLorenzo, R.~Parsey, and C.~Huang,
  ``Synthesis of patient-specific transmission data for pet attenuation
  correction for pet/mri neuroimaging using a convolutional neural network,''
  \emph{Journal of nuclear medicine}, vol.~60, no.~4, pp. 555--560, 2019.

\bibitem{ref_37}
Y.~Pan, M.~Liu, C.~Lian, T.~Zhou, Y.~Xia, and D.~Shen, ``Synthesizing missing
  pet from mri with cycle-consistent generative adversarial networks for
  alzheimer¡¯s disease diagnosis,'' in \emph{International Conference on
  Medical Image Computing and Computer-Assisted Intervention}.\hskip 1em plus
  0.5em minus 0.4em\relax Springer, 2018, pp. 455--463.

\bibitem{ref_38}
C.~R. Jack~Jr, M.~A. Bernstein, N.~C. Fox, P.~Thompson, G.~Alexander,
  D.~Harvey, B.~Borowski, P.~J. Britson, J.~L.~Whitwell, C.~Ward \emph{et~al.},
  ``The alzheimer's disease neuroimaging initiative (adni): Mri methods,''
  \emph{Journal of Magnetic Resonance Imaging: An Official Journal of the
  International Society for Magnetic Resonance in Medicine}, vol.~27, no.~4,
  pp. 685--691, 2008.

\bibitem{ref_39}
Z.~Wang, E.~P. Simoncelli, and A.~C. Bovik, ``Multiscale structural similarity
  for image quality assessment,'' in \emph{The Thrity-Seventh Asilomar
  Conference on Signals, Systems \& Computers, 2003}, vol.~2.\hskip 1em plus
  0.5em minus 0.4em\relax Ieee, 2003, pp. 1398--1402.

\bibitem{ref_40}
M.~Heusel, H.~Ramsauer, T.~Unterthiner, B.~Nessler, and S.~Hochreiter, ``Gans
  trained by a two time-scale update rule converge to a local nash
  equilibrium,'' in \emph{Advances in neural information processing systems},
  2017, pp. 6626--6637.

\bibitem{ref_41}
O.~Ronneberger, P.~Fischer, and T.~Brox, ``U-net: Convolutional networks for
  biomedical image segmentation,'' in \emph{International Conference on Medical
  image computing and computer-assisted intervention}.\hskip 1em plus 0.5em
  minus 0.4em\relax Springer, 2015, pp. 234--241.

\bibitem{ref_42}
G.~Huang, Z.~Liu, L.~Van Der~Maaten, and K.~Q. Weinberger, ``Densely connected
  convolutional networks,'' in \emph{Proceedings of the IEEE conference on
  computer vision and pattern recognition}, 2017, pp. 4700--4708.


\end{thebibliography}

\end{document}